\documentclass[sigconf,natbib=true,anonymous=false]{acmart}
\usepackage{multirow}
\AtBeginDocument{%
  \providecommand\BibTeX{{%
    \normalfont B\kern-0.5em{\scshape i\kern-0.25em b}\kern-0.8em\TeX}}}

\setcopyright{acmcopyright}
\copyrightyear{2022}
\acmYear{2022}
\setcopyright{acmcopyright}\acmConference[SIGIR '22]{Proceedings of the 45th International ACM SIGIR Conference on Research and Development in Information Retrieval}{July 11--15, 2022}{Madrid, Spain}
\acmBooktitle{Proceedings of the 45th International ACM SIGIR Conference on Research and Development in Information Retrieval (SIGIR '22), July 11--15, 2022, Madrid, Spain}
\acmPrice{15.00}
\acmDOI{10.1145/3477495.3531742}
\acmISBN{978-1-4503-8732-3/22/07}

\begin{document}
\fancyhead{}

\title{Biographical: A Semi-Supervised Relation Extraction Dataset}

\author{Alistair Plum}
\orcid{0000-0003-0977-3467}
\affiliation{RGCL,\\
 \institution{University of Wolverhampton, UK}
 \country{}}
 \email{a.j.plum@wlv.ac.uk}
 
 \author{Tharindu Ranasinghe}
  \orcid{0000-0003-3207-3821} 
\affiliation{RGCL,\\
 \institution{University of Wolverhampton, UK}
 \country{}}
\email{tharindu.ranasinghe@wlv.ac.uk}

 \author{Spencer Jones}
  \orcid{0000-0001-6241-5971}
\affiliation{History, Politics and War Studies,\\
 \institution{University of Wolverhampton, UK}
 \country{}}
\email{spencer.jones@wlv.ac.uk}

\author{Constantin Or\u{a}san}
\orcid{0000-0003-2067-8890}
\affiliation{Centre for Translation Studies, \\
 \institution{University of Surrey, UK}
 \country{}}
\email{c.orasan@surrey.ac.uk}

\author{Ruslan Mitkov}
\orcid{0000-0003-2522-066X}
\affiliation{RGCL,\\
 \institution{University of Wolverhampton, UK}
 \country{}}
\email{r.mitkov@wlv.ac.uk}

\renewcommand{\shortauthors}{Plum et al.}

\begin{abstract}
Extracting biographical information from online documents is a popular research topic among the information extraction (IE) community. Various natural language processing (NLP) techniques such as text classification, text summarisation and relation extraction are commonly used to achieve this. Among these techniques, RE is the most common since it can be directly used to build biographical knowledge graphs. RE is usually framed as a supervised machine learning (ML) problem, where ML models are trained on annotated datasets. However, there are few  annotated datasets for RE since the annotation process can be costly and time-consuming. To address this, we developed \textit{Biographical}, the first semi-supervised dataset for RE. The dataset, which is aimed towards digital humanities (DH) and historical research, is automatically compiled by aligning sentences from Wikipedia articles with matching structured data from sources including Pantheon and Wikidata. By exploiting the structure of Wikipedia articles and robust named entity recognition (NER), we match information with relatively high precision in order to compile annotated relation pairs for ten different relations that are important in the DH domain. Furthermore, we demonstrate the effectiveness of the dataset by training a state-of-the-art neural model to classify relation pairs, and evaluate it on a manually annotated gold standard set. \textit{Biographical} is primarily aimed at training neural models for RE within the domain of digital humanities and history, but as we discuss at the end of this paper, it can be useful for other purposes as well.

% We believe that \textit{Biographical} would pave the way for semi-supervised RE methods.

    % In this paper we present \textit{Biographical}, a semi-supervised dataset for relation extraction, as well as the accompanying compilation process. The dataset, primarily aimed at training neural models for relation extraction within the domain of digital humanities and history, is automatically compiled by aligning sentences from Wikipedia articles with matching structured data from sources, including Pantheon and Wikidata. By exploiting the structure of Wikipedia articles and robust named entity recognition, we are able to match information with relative ease and precision in order to compile annotated relation pairs. The process description includes for ten different relations, which require slightly different processing approaches. 

    % Experiments carried out include different text processing approaches in order to determine a good balance between amount and variety of data. We also demonstrate the effectiveness of the dataset by training two neural models to classify relation pairs, and present an evaluation using a gold-standard dataset, which was manually annotated for this project. Further research presented here concerns adding further relations to the dataset in the future, as well as the implications of such a dataset in the areas of digital humanities and history.
\end{abstract}

%\vspace{-6mm}

\begin{CCSXML}
<ccs2012>
   <concept>
       <concept_id>10010147.10010178.10010179.10003352</concept_id>
       <concept_desc>Computing methodologies~Information extraction</concept_desc>
       <concept_significance>500</concept_significance>
       </concept>
   <concept>
       <concept_id>10010147.10010178.10010179.10010186</concept_id>
       <concept_desc>Computing methodologies~Language resources</concept_desc>
       <concept_significance>500</concept_significance>
       </concept>
 </ccs2012>
\end{CCSXML}

\ccsdesc[500]{Computing methodologies~Information extraction}
\ccsdesc[500]{Computing methodologies~Language resources}

\keywords{Biographical Information Extraction, Relation Extraction, Transformers}

%\vspace{-2mm}

\maketitle

\section{Introduction}
As web technology continues to thrive, documents containing biographical information are continuously generated and published online in large numbers \cite{10.1145/3445965}. These online documents contain essential facts or events related to the life of well-known and lesser-known individuals, which can be used to populate structured biographical databases \cite{10.1145/3451471.3451506, 10.1145/3241741}. These databases are capable of supporting many interesting studies in humanities, and related areas \cite{zhang-etal-2017-position} as we describe in Section \ref{sec:discussion}. However, manually extracting information from a massive document collection is impossible, given the amount of information available online. Therefore, NLP methods can be used to process these documents automatically. 

Previous studies have used many NLP techniques including text classification  \cite{Aprosio2015, Hogue2014}, named entity recognition (NER) \cite{Jiang2012} and summarisation \cite{Zhou2004} to perform biographical information extraction, which we describe thoroughly in Section \ref{sec:related_work}. However, a major weakness in these studies is that they can not be used directly to populate a database. Instead, they need to be combined with other NLP techniques to extract the structured information required for databases. A different approach, which we employ in this study, is to design biographical information extraction as a relation extraction (RE) task. 

RE is the task of extracting semantic relationships between entities from a document, which can in turn be used to populate a database with relational facts contained in a piece of text. Consider the following two text pieces on two different people. 

\vspace{2mm}
\noindent\fbox{%
    \parbox{\columnwidth}{%
        \textit{\textbf{Text 1:}} William Shakespeare was born and raised in Warwickshire. At the age of 18, he married Anne Hathaway, with whom he had three children: Susanna Hall and twins Hamnet Shakespeare and Judith Quiney.
    }%
}

\vspace{2mm}
\noindent\fbox{%
    \parbox{\columnwidth}{%
        \textit{\textbf{Text 2:}} Henry Baynton (23 September 1892 in Warwickshire – 2 January 1951 in London) was a British Shakespearean actor of the early 20th century.
    }%
}

\vspace{2mm}

\begin{figure}[!ht]%!ht
\centering
\scalebox{0.58}{\includegraphics{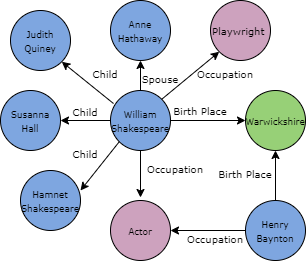}}
\caption{Example Knowledge Graph}
\label{fig:knowledge_graph}
\end{figure}

For the texts shown above, the RE model can extract triples, which can be represented as edges in a knowledge graph, such as <William Shakespeare, \textit{Spouse}, Anne Hathaway>. Table \ref{tab:examples} shows some of the relationship triples that can be extracted from the above two text pieces. Combining such triples, a system can produce a knowledge graph of relational facts between persons, occupations, and locations in the text. A knowledge graph derived from the relationships in Table \ref{tab:examples} is shown in Figure \ref{fig:knowledge_graph}.

\begin{table}[ht]%ht
\scalebox{1.00}{
\begin{tabular}{|l|l|l|}
\hline
 \textbf{Object} & \textbf{Relation} &  \textbf{Object}  \\ 
 \hline
William Shakespeare & \textit{Birth Place}  &  Warwickshire  \\
William Shakespeare & \textit{Spouse}  &  Anne Hathaway  \\
William Shakespeare & \textit{Child}  &   Susanna Hall  \\
William Shakespeare & \textit{Child}  &  Hamnet Shakespeare  \\
William Shakespeare & \textit{Child}  &  Judith Quiney  \\
William Shakespeare & \textit{Occupation}  &  Actor  \\
William Shakespeare & \textit{Occupation}  &  PlayWright  \\
Henry Baynton & \textit{Occupation}  &  Actor  \\
Henry Baynton & \textit{Birth Place}  &  Warwickshire  \\ \hline
\end{tabular}
}
\caption{Example Biographical Relationship Triples}
\label{tab:examples}
\end{table}

%%%%%%%%%%%%%%%%%%%%%%%%%%%%%%%%%%%%%%%%%%%%%%%%%%%%% original placement
% \begin{figure}[!ht]
% \centering
% \scalebox{0.58}{\includegraphics{images/knowledge_graph_1.png}}
% \caption{Example Knowledge Graph}
% \label{fig:knowledge_graph}
% \end{figure}

Knowledge graphs are commonly used by companies to provide information to end-users and understand relationships between various types of entities. Several machine learning models including recurrent neural networks (RNN) \cite{gormley-etal-2015-improved, xiao-liu-2016-semantic}, convolutional neural networks (CNN) \cite{zeng-etal-2014-relation, shen-huang-2016-attention}, graph neural networks (GNN) \cite{baldini-soares-etal-2019-matching, Xue_Sun_Zhang_Chng_2021} and transformers \cite{Nayak_Ng_2020, joshi-etal-2020-spanbert} have been proposed to automatically extract relationships from texts. These machine learning models use a supervised paradigm where the models require a dataset similar to Table \ref{tab:examples} to train. Therefore, the NLP community has a growing interest in producing datasets capable of training machine learning models to perform RE. Several datasets in this area, such as NYT24 \cite{hoffmann-etal-2011-knowledge}, and TACRED \cite{zhang-etal-2017-position} have been released for this purpose. However, all of these datasets are manually annotated, which makes it difficult to expand RE to different genres and languages. In this paper, we propose a novel approach for producing RE datasets that is semi-supervised and can be expanded easily to other domains and languages. As far as we know, an approach such as this has not yet been proposed. We develop the first dataset of this kind and evaluate its usefulness. If the approach does prove to be useful, it will significantly reduce the burden on the manual annotation process, as well as language and domain-specific expertise. 

The main contributions of this paper are the following:

\begin{enumerate}
    \item We introduce Biographical; the first and the largest dataset for biographical RE built in a semi-supervised manner with ten relationship categories. We also produce a manually annotated subset that can be used for evaluation\footnote{The dataset is available at \url{https://plumaj.github.io/biographical/}}.

    \item We evaluate four machine learning models to perform biographical RE, based on state-of-the-art transformer models such as BERT \cite{devlin2019bert}.

    \item We provide important resources to the community: the dataset, the code, and the pre-trained models are made available to everyone interested in working on biographical RE using the same methodology.
\end{enumerate}

The rest of the paper is structured as follows. Section \ref{sec:related_work} presents an overview of related work. Section \ref{sec:method} describes the data compilation process involved in this study. In Section \ref{sec:exp_eval} we explain the experiments carried out, as well as an evaluation of the experiments. Finally, the paper outlines an intended future study and provides conclusions.

\section{Related Work}
\label{sec:related_work}
Extracting biographical information from documents is a popular research area in the NLP community. Most of these studies use different NLP techniques on open and free resources such as Wikipedia. 

Text classification is one of the first NLP techniques used to extract biographical information. \citet{Biadsy2008} used an unsupervised sentence classification framework to extract biographies from Wikipedia articles. In more recent work, \citet{Aprosio2015} have trained various machine learning classifiers to detect biographical sections in Wikipedia texts using a supervised approach. In a different work, \citet{Hogue2014} use Wikipedia page traffic data to determine sentences of importance in Wikipedia articles.

Text summarisation is another popular NLP technique that has been used to extract biographical information. \citet{Biadsy2008} use Wikipedia articles together with the TDT4 news corpus\footnote{\url{https://catalog.ldc.upenn.edu/LDC2005S11}} to train an unsupervised multi-document summariser for biographical information. They used a support vector machine model and achieved state-of-the-art performance at the time on the DUC2004 dataset \cite{DUC2004}. The approach is based on the one proposed by \citet{Zhou2004} who similarly used Wikipedia data to develop a system for summarisation using a Naive Bayes architecture. \citet{Chisholm2017} combine Wikipedia text and Wikidata information to generate one-sentence summaries from structured biographical information. First, the approach identifies potential biographical candidates from Wikidata, then learns to generate the short summaries by mapping structured information to the first sentence of the matching article in Wikipedia. Thus follows the mostly standardised pattern of the first sentence of a Wikipedia article containing most of the relevant information about a person. 

However, none of these approaches can be used directly to create a knowledge graph. Therefore, more recent work in biographical information extraction has modelled the task as a RE problem. Several ML models have been developed to perform RE. Early approaches for RE were based on traditional machine learning models such as support vector machines \cite{10.5555}, and decision trees \cite{Singhal2016bm}. But with the introduction of word embeddings and the success of neural network architectures in different areas, the NLP community has used a wide range of neural network architectures for the RE task. \citet{zeng-etal-2014-relation} have used a CNN architecture and a synonym dictionary to integrate semantic knowledge into the neural network. In a different approach, \citet{zeng-etal-2014-relation} use lexical features with the word embeddings \cite{turian-etal-2010-word} fed into a CNN to perform RE. RNNs have also been popularly used in RE. \citet{miwa-bansal-2016-end} utilised a Tree Long Short-Term Memory (LSTM) network to perform RE. \citet{zhou-etal-2016-attention}  used an attention-based bi-directional LSTM network on the SemEval-2010 relation classification task \cite{hendrickx-etal-2010-semeval} and show that it provides good results. The current state-of-the-art in RE, also used for this research, is based on neural transformers \cite{baldini-soares-etal-2019-matching}. These transformer models are trained using a language modelling task such as masked language modelling or next sentence prediction and then have been used to perform RE as a downstream NLP task. Results on recent RE datasets show that transformers outperform the previous architectures based on RNNs and CNNs \cite{8983370,baldini-soares-etal-2019-matching}. 

All the ML models for RE mentioned above follow a supervised paradigm where an annotated dataset is required to train the ML model. The most common datasets used for this are NYT24 \cite{hoffmann-etal-2011-knowledge}, NYT29 \cite{10.1007/978-3-642-15939-8_10} and TACRED \cite{zhang-etal-2017-position}. All these datasets have been created using manual annotation. As we mentioned before, since the annotation process is expensive, these datasets are limited in size. For example, TACRED, the largest RE dataset,  has only 106,264 instances. This can prove not enough to train data-driven methods, especially those based on neural networks. Furthermore, the manual annotation process limits the expansion of RE research to different domains and languages. To address this problem, we propose a semi-supervised approach to create RE datasets using a similar approach to \citet{Chisholm2017} which we describe in the next section.

% Using Wikipedia to detect, process and extract biographical information is by no means novel. \cite{Aprosio2015} have trained various machine learning classifiers to detect biographical sections in Wikipedia texts using a supervised approach. 

%  An approach with the same aim but different method is presented by \citet{Hogue2014}, who use Wikipedia page traffic data to determine sentences of importance in Wikipedia articles.

% Similar to the proposed methodology of the present approach, \citet{Chisholm2017} combine Wikipedia text and Wikidata information to generate one sentences summaries from structured biographical information. The approach first identifies potential biographical candidates from Wikidata, then learns to generate the short summaries by mapping structured information to the first sentence of the matching article in Wikipedia. Thus follows the mostly standardised pattern of the first sentence of a Wikipedia article containing most of the relevant information about a person.

\begin{figure}[!ht]%!ht
\centering
\scalebox{0.6}{\includegraphics{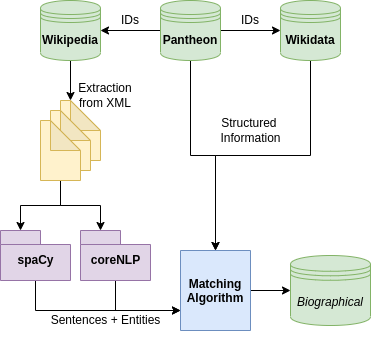}}
\caption{System Architecture}
\label{fig:architecture}
\end{figure}

\section{Data Compilation}
\label{sec:method}
The data compilation process is divided into two steps. The first step involves the selection of our data sources, which are one of the most fundamental aspects of the approach (Section \ref{sec:method_data}). Our approach requires a source of textual data and a source of structured information that is related to the textual data. The second step concerns the processing of the different data sources, as well as matching operations that allow for the automatic labelling process (Section \ref{sec:method_labels}). These steps lead to the final dataset consisting of sentences, marked entities and their respective relation.

%%%%%%%%%%%%%%%%%%%%%%%%%%%%%%%%%%%%%%%%%%%%%%%% original placement
% \begin{figure}[!ht]%!ht
% \centering
% \scalebox{0.62}{\includegraphics{images/arch_col.png}}
% \caption{System Architecture}
% \label{fig:architecture}
% \end{figure}

%It is also important to note the restrictions set for this approach, as this is a key aspect of the approach described here. We do not pursue an open information extraction oriented approach, where the goal would be to extract as much information as possible, including any kind of relation. However, with the semi-supervised approach at hand this would not be compatible, since there would be too many variables. We therefore explicitly target specific structured information in sentences and limit the relations we extract. In spite of the restrictions, this does not mean that further relations could be added to the dataset at a later point.

\subsection{Data Sources}
\label{sec:method_data}
Our semi-supervised approach combines data from three different sources: Wikipedia\footnote{\url{https://www.wikipedia.org/}}, Wikidata\footnote{\url{https://www.wikidata.org/wiki/Wikidata:Main_Page}} and Pantheon\footnote{\url{https://pantheon.world/}} \cite{Yu2016}. Wikipedia serves as the main source of textual data, in the form of sentences taken from specific articles. Pantheon and Wikidata serve as our sources of structured information. We also use Pantheon to select our initial set of biographical articles from Wikipedia. We target specific biographical articles in Wikipedia that are confirmed by the Pantheon dataset. Next, we iterate over the sentences of each article and tag the named entities, including locations and dates, using spaCy\footnote{\url{https://spacy.io/}} and Stanford CoreNLP\footnote{\url{https://stanfordnlp.github.io/CoreNLP/}}. Finally, we augment the structured data from the Pantheon dataset with information from Wikidata. This expanded dataset is matched to sentences in Wikipedia, allowing us to label each sentence according to the type of relation. We discuss each of the data sources in more detail in the following sections.

\subsubsection{Wikipedia}
Wikipedia is a free, online encyclopedia that contains a large amount of information about people, and as such, serves as the backbone of our approach. It is a vast resource of textual data, that is linked to a number of different projects that relay the contained information in a structured way. The next steps in our approach will focus on connecting the structured data with the textual data.

% This is an ideal situation for our approach, since all we have to do is find ways of connecting the structured data with the textual data.

For processing Wikipedia textual data, we follow a previously established workflow \cite{plum2022} which has proved to be efficient. We work with Wikipedia database backup dumps, which are an exact copy of all Wikipedia articles of a given language at a specific point in time. We use the \textit{enwiki-20190420} dump, which corresponds to the content of English Wikipedia on 20th of April 2019. Once downloaded, we extract articles corresponding to the entries in the Pantheon dataset, which is done via the Wikipedia IDs. Extracting the text can be a complex task in itself, since the structure of the XML file is not uniform, as well as including certain XML parts that have to be expanded. Since the extraction of text from Wikipedia is not our main goal and could warrant a separate project, we use an existing tool for the extraction process. The wikiextractor\footnote{\url{https://github.com/attardi/wikiextractor}} package for Python converts articles to plain text. We observed some extraction problems, such as XML-tag artefacts, mismatched quotation marks, and incomplete or illegible sentences, which we remove at the processing stage with regular expressions.

\subsubsection{Pantheon}
In order to determine which articles in Wikipedia are biographical, i.e. containing information that pertains to a person, we use the Pantheon dataset \cite{Yu2016}. According to its creators, \emph{"Pantheon [is] focused on biographies with a presence in 15 different languages in Wikipedia"} and consists of roughly 85,000 entries. While it was initially created mostly by hand, its later iterations have used a classifier to determine and extract further entries. One particular characteristic of this dataset is that each article has to contain unambiguous links to the respective Wikipedia and Wikidata pages. This allows us to identify which articles from Wikipedia contain the relevant information. While this could be done just using Wikidata, Pantheon has been (at least partly) manually verified. Because Pantheon only includes persons whose articles are available in 15 different languages, this ensures that a person is somewhat well-known, in turn making a longer Wikipedia article more likely.

In addition, each entry includes basic information, which we match to sentences from the corresponding Wikipedia articles. This mainly includes information such as dates of birth and death, places of birth and death, and main occupation. The included information allows us to label the \textit{birthdate, deathdate, birthplace, deathplace} and \textit{occupation} relations, while also allowing us to confirm the name of a person. As these relations are only half of the relations we target, we use the included Wikidata ID to obtain the other half of the relations (introduced next).

\subsubsection{Wikidata}
Wikidata is described as a \emph{"free, collaborative, multilingual, secondary database" that "provides support for Wikipedia [...]"} \cite{wikidata2014}. Wikidata ties in well with the two other sources of data that we use. Since it provides most of the information from a Wikipedia page (and often more) in a structured format, we use it to augment the Pantheon dataset. Since the Pantheon dataset provides distinct identifiers for Wikipedia and Wikidata, selecting the correct entity is a straight-forward task. Using the corresponding entries, we add the \textit{educatedAt, ofParent, sibling} and \textit{hasChild} relations, as well as \textit{other}. In the case of the last relation, we use this to categorise any relation that is not explicitly targeted here and make sure that the information matched is not part of any of the nine other relations.

\subsection{Automatic Labelling}
\label{sec:method_labels}
The next step in the approach is the automatic labelling of sentences. Once we have extracted the text of each Wikipedia article, we begin processing the texts, using spaCy NER to tag persons, locations, organisations, dates, as well as Stanford CoreNLP Entity information to tag occupations in each article. It should be noted that we run spaCy at runtime, but we carried out one full annotation run with Stanford CoreNLP on all articles, which we store and subsequently only access. This is because we found Stanford CoreNLP too slow for multiple runs. 

Each sentence of an article is processed in order to determine whether it is about the main person of the article. To accomplish this, the script matches the name with the person tags in the sentence, and also allows some substring matches, such as first and last name excluding any other titles, or last name only. If a match is found, the sentence is regarded as containing some information about that person. This is ensured because the sentence is taken from that person's article and it includes that person's name.

After a positive match is made within a sentence, we check the other tagged entities in the sentence (locations, organisations, dates and occupations) against the information provided by the Pantheon dataset and respective Wikidata entry. Each matched pair, for instance a \textit{name} and a \textit{location}, is then marked with $<$eN$>$ (begin) and $<$/eN$>$ (end) tags, where N is either 1 or 2, depending on the position of the entity (i.e. first or last). This is followed by the respective relation tag. The following text box shows an example of this. We estimate that this approach could be extended to all relations where it would be possible match the information in a sentence in this way. 

\vspace{2mm}
\noindent\fbox{%
    \parbox{\columnwidth}{%
        \textit{\textbf{Text 1:}} \textbf{<e1>}William Shakespeare\textbf{</e1>} was born and raised in \textbf{<e2>}Warwickshire\textbf{</e2>}.
    }%
}
\vspace{2mm}

% We split the final dataset into a training set (80\%), a development set (10\%) and a test set (10\%). Additionally, we also extract 100 sentences per relation, which me manually annotated and refer to as the gold set. More details about this process are described in Section \ref{sec:exp_manual}.

We hypothesise that this simple combination of named entity tagging and string matching works because of the controlled circumstances, which were mentioned at the beginning of this section. We only allow matches involving the person who is the main subject of an article, ensuring that statements made in sentences are most likely to be about this person. This may sound quite obvious at first. However, sentences taken from articles at random, matching random people, do not necessarily contain statements about that person. If the subject of the Wikipedia article is a certain person, most, if not all, statements made mentioning that person are likely to directly relate to that person. 

Another control mechanism involves the structure of Wikipedia. Often we find a number of opening paragraphs containing the most important information about a person (or other entity). First mentions of certain facts are likely to be the main information, such as the first date mentioned usually being the date of birth, first mentioned locations being the places of death and/or birth, job titles usually the corresponding (and main) occupation of the person and so on. It should be mentioned, however, that this structure can cause problems, as will be elaborated on in section \ref{sec:exp_manual}.

It is important to note that not every relation is always found for every entity. We therefore tried different processing approaches for the textual data, detailed in Section \ref{sec:exp_proc}. A breakdown of the number of relations per set is presented in Section \ref{sec:exp_manual}. Each relation also requires slightly different handling depending on the type of information. Tasks include date normalisation, partial matching for occupations, and exact location name matching. Exact details are presented in the following sections. 

\subsubsection{Date-based Relations}
This set of relations includes \textit{birthdate} (date of birth) and \textit{deathdate} (date of death). In order to match these relations, the system checks for a \textit{DATE} entity in the sentence, which is normalised to \textit{YYYY-MM-DD} format. We use the dateparser\footnote{\url{https://dateparser.readthedocs.io/en/latest/}} package and use the date of processing as a relative date (for rare cases such as \textit{tomorrow} or \textit{today}). Furthermore, we use the first match for both relations, discarding subsequent matches. This mode of processing aligns with our restrictive approach, which assumes most pertinent information to mentioned towards the beginning of a Wikipedia article, rather than towards the end. 

\subsubsection{Name-based Relations}
This set of relations includes \textit{ofParent, sibling} and \textit{hasChild}, as well as \textit{educatedAt} (the place of education). For these name-based relations, the system checks a sentence for \textit{PER} and \textit{ORG} tags. It is ensured that only full matches are accepted, even though it may seem favourable to accept partial matches, at least for anything concerning persons. This is because with persons, it can be reasonable to allow just the first or last name to match. However, we found during the manual annotation process (Section \ref{sec:exp_manual}) that too many false matches occurred, caused by different persons having the same name.

\subsubsection{Entity Information Relations}
Only the \textit{occupation} relation is included in this group. Since spaCy's NER capabilites do not include any tags such as \textit{title} or \textit{job}, we opted to use Stanford CoreNLP's entity information processing to add this relation. We could have trained the spaCy model to include a new entity type for this step. In the end, we used CoreNLP as we felt training a new relation could potentially introduce another layer of errors. 

The system lookup for this relation functions in a similar way to the previous set of relations, only that instead the CoreNLP information is accessed for matching. As mentioned, we run the initial CoreNLP processing separately due to the increased run time.  Again, we only allow the complete first match to be annotated. Potentially, this relation set could be extended by using further occupation information from Wikidata, which in most cases lists a number of different occupations for a person, rather than the one main occupation listed in Pantheon.

\subsubsection{Other Relations}
% other is anything that hasn't matched at all (so cases that could match but ultimately fail are not included)
This class of relations, labelled as \textit{other} in the dataset, is used for all other relations. It is essentially the zero class, that is labelled when all other lookups in a sentence have failed. The \textit{other} label is then applied to an entity pair that does not appear to be part of any of the other nine relations matched. Since we obtain more sentence from this class than all the other nine combined, we randomly select sentences and balance according to the total number of all other sentences containing relations. We balance the \textit{other} relation class to make it equivalent in size with the remaining nine relations combined. 

If in future more relations are added to the dataset, it would be vital to ensure that these \textit{Other} labelled sentences do not contain the new relations, since they could conceivably be anything.

\section{Experiments}
\label{sec:exp_eval}
We carried out multiple experiments to estimate the quality and usefulness of this dataset. First, we examined the effects of different processing approaches for the article texts. Next, we manually annotated a small sub-set of sentences to pinpoint potential problems and to create a gold-standard set for evaluation purposes. After re-running the compilation process, taking into account certain observations and minor processing improvements after manual annotation, we trained a number of state-of-the-art ML models using the training datasets, and evaluate the performance using the gold set.

\subsection{Labelling Approaches}
\label{sec:exp_proc}
%different sets for variety of text, examples
For the process of automatically labelling each entity pair with a corresponding relation, we work at the document and sentence levels of a relevant Wikipedia article. At the document level we carry out all the NLP processing, such as NER, and then split the article into its sentences, to process each sentence. However, we wanted to assess the effect of two further approaches of processing the articles: First, we wanted to see how well co-reference resolution performs on the Wikipedia texts, and whether it would yield more annotated sentences (Section \ref{sec:exp_coref}). Next, we looked into addressing sentence diversity, by implementing an approach that skips the first sentence of an article (Section \ref{sec:exp_skip}).

\subsubsection{Coref Set}
\label{sec:exp_coref}
We hypothesise that replacing co-referential entity mentions will allow the matching algorithm to find more matches overall. This would be due to the fact that more names would be matched because of the increased presence. Detecting more names could then potentially lead to more relation matches overall. For this, we used spaCy's built-in co-reference resolution capabilities to automatically replace entity mentions with the most probable entity. The matching step is carried out using the text where all the entities have been replaced. 

Table \ref{tab:no_rels} shows the number of relations found across each of the sets we compiled: normal, coref, which is described here, and skip, which is described in the next section. The last line of the table shows the total number of relations found per set.

\begin{table}[!ht]
\begin{tabular}{|l|c|c|c|}
\hline
             & \textbf{normal} & \textbf{coref} & \textbf{skip}   \\ \hline
birthdate    & \textbf{52,083} & 48,004           & 45,366  \\
birthplace & \textbf{50,396} & 46,552           & 19,746  \\
deathdate    & \textbf{17,376} & 14,505           & 87,93   \\
deathplace & 19,055          & \textbf{20,444}  & 11,202  \\
occupation   & \textbf{41,469} & \textbf{41,469}  & 17,642  \\
ofParent       & 6,503           & \textbf{10,301}  & 6,022   \\
educatedAt   & 5,738           & \textbf{9,430}   & 5,694   \\
hasChild        & 2,343           & \textbf{4,042}   & 2,215   \\
sibling      & 2,189           & \textbf{3,618}   & 2,098   \\ 
other        & 197,952         & \textbf{199,165} & 119,578 \\ \hline
\textbf{Total}       & 395,104        & \textbf{397,530}            & 238,356 \\ \hline
\end{tabular}
\caption{Number of relations in each set: \textit{normal} uses the normal processing method, \textit{coref} uses automatic coreference resolution and \textit{skip} skips the first sentence of each article.}
\label{tab:no_rels}
\end{table}

If we compare the overall counts of the relations of the normal and coref sets, we observe a small increase. However, looking at the counts of the different relation types, we see that it is not a simple increase across the board. In fact, we see fewer matches in certain cases. Upon further inspection, we found that this was mainly due to the automatic replacement process producing illegible sentences through incorrect replacements. The two main problems were entities that are scrambled and sentences being unintelligible because every single entity mention was replaced with one main entity, that was often also too long. 

The main problems are demonstrated in the two examples below. In the first sentence, an entity has been replaced many times, including an opening bracket. Cases like these were observed frequently, and with more characters added. These cases introduced matching errors in the set. In the second example, we see a nested replacement, which similarly causes matching problems.

\vspace{2mm}
\noindent\fbox{%
    \parbox{\columnwidth}{%
        \textit{\textbf{Replaced:}} Born in <e1>Évreux</e1>, Eure, a great fan of \textbf{Paris Saint-Germain Paris Saint-Germain} since   \textbf{<e2>Bernard Mendy</e2> (} childhood,  \textbf{ Bernard Mendy (} achieved \textbf{ Bernard Mendy (} ambitions in 2000 when   \textbf{Bernard Mendy (} joined PSG from SM Caen. \\
        \textit{\textbf{Original:}} Born in <e1>Évreux</e1>, Eure, a great fan of Paris Saint-Germain since his childhood, he achieved his ambitions in 2000 when he joined PSG from SM Caen. 
    }%
}

\vspace{2mm}

\noindent\fbox{%
    \parbox{\columnwidth}{%
        \textit{\textbf{Replaced:}} The hundreds of volumes contained \textbf{Queen Victoria's Queen <e1>Victoria</e1>'s's} personal views of [...] \\
        \textit{\textbf{Original:}} The hundreds of volumes contained \textbf{Queen Victoria's} personal views of [...]
    }%
}

\vspace{2mm}

To understand better why this approach does not work well, we carried out a manual annotation of 100 randomly selected sentences per relation from this set, which is described in Section \ref{sec:exp_manual}. We also trained a neural model using this dataset, the evaluation of this is detailed in Section \ref{sec:neural_eval}.

\subsubsection{Skip Set}
\label{sec:exp_skip}
The \textit{skip set} was compiled to study the effects of leaving out the first sentence of an article from Wikipedia. One problem with using Wikipedia texts stems from the first sentence of an article, or rather the structure of the first sentence of an article, is seen in the following example.

\vspace{2mm}
\noindent\fbox{%
    \parbox{\columnwidth}{%
        William Shakespeare (bapt. 26 April 1564 – 23 April 1616) was an English playwright, poet and actor, widely regarded as the greatest writer in the English language and the world's greatest dramatist.
    }%
}
\vspace{2mm}

We see that the date of birth (and death) occur within parentheses after the name, in addition to the fact that the sentence usually contains a large amount of summarised information. This type of sentence structure (and content) is not only extremely frequent, but also quite specific to Wikipedia, suggesting that unnatural behaviour could be learned by a machine learning model. This was observed by \citet{Chisholm2017} who exploited this for their benefit. However, for this approach, we wanted to achieve as many natural matches as we could. Therefore, we compiled a dataset that follows the previously described methodology, but skips the first sentence of each article. The hypothesis is that this forces more matches elsewhere in the article, where more natural sentences occur.

Table \ref{tab:no_rels} shows the total and individual counts for each relation, as referred to previously. We see that overall, the skip set has much fewer matches than the other two sets, and it never has the highest number of individual counts in any category, although the numbers are comparable in some categories to the normal set. Regardless, some of the generally larger categories, such as \textit{birthplace} and \textit{birthdate} are significantly smaller than the other two sets, generally pointing towards the fact that the identification is successful, as this information is extremely common in the first sentence. It is not always certain that this information will appear later on in an article, therefore leading to a smaller number of matches. 

As with the previous set, we present a manual evaluation of 100 randomly selected sentences per relation from this set in Section \ref{sec:exp_manual}, and the results of a trained neural model using this dataset in Section \ref{sec:neural_eval}.

\subsection{Manual Annotation}
\label{sec:exp_manual}
We assessed the quality of our semi-supervised datasets before using it to train machine learning models by means of manual annotation. This was important in order to find areas where the approach fails to match data accurately, where processing methods do not work, and any other similar problems. In addition, we needed a gold standard test set for benchmarking our neural models. 

As pointed out in previous sections, we extracted 100 sentences per relation across the three datasets, equalling 3000 sentences in total that we manually annotated and refer to as the gold set. The data was annotated by two persons, one native English speaker and one non-native but fluent English speaker, both postgraduate students. For each sentence, the task was to look at the relation assigned by our matching algorithm and add the correct relation if it had been labelled incorrectly. We used one of the nine indicative labels where appropriate, and the \textit{other} label if a different relation was expressed. Our annotation guideline was that a human should understand by reading the sentence which relation is expressed, regardless of prior knowledge. This is demonstrated by the following examples. 

The first example shows a sentence that clearly mentions the \textit{occupation} E2 of the entity E1. The second example shows an implicit relation. Although it is not directly stated, the word \textit{orphaned} in relation to entity E2 with the statement that E1 \textit{died}, implies that E1 is the \textit{parent} of E2. In the final example, the algorithm labels the sentence as expressing the \textit{parent} relation between the two entities. Although this may indeed be the case, and the annotator may have prior knowledge of this, or it has been expressed in a different sentence, it is not clearly stated in this sentence.

\vspace{2mm}
\noindent\fbox{%
    \parbox{\columnwidth}{%
        \textit{\textbf{Explicit:}} <e1>\textbf{Renate Künast}</e1> (born 15 December 1955) \textbf{is a} German <e2>\textbf{politician}</e2> of Bündnis 90/Die Grünen.
    }%
}

\vspace{2mm}
\noindent\fbox{%
    \parbox{\columnwidth}{%
        \textit{\textbf{Implicit:}} A few months later <e1>\textbf{Apollo Korzeniowski}</e1> \textbf{died}, leaving <e2>\textbf{Conrad}</e2> \textbf{orphaned} at the age of eleven.
    }%
}

\vspace{2mm}

\noindent\fbox{%
    \parbox{\columnwidth}{%
        \textit{\textbf{Unclear:}} Thus, <e1>\textbf{Janaka}</e1> tries to find the best husband for <e2>\textbf{Sita}</e2>.
    }%
}

\vspace{2mm}

The Cohen's Kappa for the inter-annotator agreement is 0.908 which indicates a very high agreement between our annotators. The annotations allowed us to make a number of observations. First, we notice that two very similar relations work very differently. While \textit{birthplace} works extremely well across sets, \textit{deathplace} does not. Upon further examination, we found that the first mention of the place where someone died often was also the place where a person lived. In future, cases like these may warrant a different approach to processing by our algorithm, but for now we leave it unchanged. Second, we observed that many relations in the coref set were incoherent and probably incorrect, due to imprecise replacements by the coreference resolution algorithm. 

While Wikidata as a source does work quite well, categories can sometimes be ambiguous, such as the \textit{educatedAt} and \textit{parent} relations. Here, we observed that the Wikidata entries contained information at odds with our interpretation of the type of entry, such as \textit{educatedAt} containing a University that is the place of work, or \textit{parent} containing a person that the target entry is a parent \textit{of} rather \textit{has}. Since this did not occur often in our manual evaluation, we did not implement a strategy to solve this problem.

Finally, we found a number of simple processing errors that we solved by improving our regular expressions for text cleaning. We also adjusted the matching procedure for the \textit{occupation} relation, to avoid matches where the occupation mentioned belonged to a different entity. This leads to a slightly smaller number of relations overall, with a detailed overview shown in Table \ref{tab:no_rels_improv}.

\begin{table}[]%ht
\begin{tabular}{|l|l|l|l|}
\hline
             & \textbf{normal} & \textbf{coref} & \textbf{skip}   \\ \hline
birthdate    & \textbf{51,524} & 47,977           & 45,211  \\
birthplace & \textbf{50,226} & 46,551           & 17,537  \\
deathdate    & \textbf{17,197} & 14,500           & 5,925   \\
deathplace & 18,944          & \textbf{20,430}  & 10,790  \\
occupation   & \textbf{18,114} & \textbf{18,111}  & 8,716   \\
ofParent       & 6,352           & \textbf{10,291}  & 5,596   \\
educatedAt   & 5,639           & \textbf{9,415}   & 3,858   \\
hasChild        & 2,209           & \textbf{4,053}   & 2,123   \\
sibling      & 2,083           & \textbf{3,601}   & 1,997   \\
Other        & 173,969         & \textbf{175,916} & 103,248 \\ \hline
\textbf{Total}       & 346,257        & \textbf{350,845}            & 205,001 \\ \hline
\end{tabular}
\caption{Relations per Set after Processing Improvements}
\label{tab:no_rels_improv}
\end{table}

Overall, we have formed the following impressions for each set. The normal approach works well, while not offering a very diverse set of sentences. As alluded to earlier, it is clear that this approach matches mainly the standard Wikipedia first sentence, as described in previous sections. The coref set, while seemingly the largest set, must also include the most unusable sentences and bad examples. During the course of evaluating the sentences we found this set to be imprecise, not explicit and difficult to understand due to bad replacements. Finally, we found the skip set to be very mixed in terms of success. While for some relations it seems that none of the matching has returned usable results, other relations seem to have worked very well, offering in addition a wide variety of different sentences demonstrating the desired effects.

In order to determine the performance of the matching algorithm, we present the evaluation metrics for the gold set. For this, we compared the labels produced by the automatic matching algorithm to our manually produced labels. 
We removed 100 sentences from the gold set that contained processing errors caused by conversion to plain text, automatic replacement of coreferences and spaCy tagging errors. Since these would all have been annotated as "Other", we decided to remove these sentences since they could have caused an imbalanced test set. Table \ref{tab:gold_eval} shows the results of the evaluation for each set. We observe that most of the matches found are correct, indicated by high precision and recall scores. However, the problem with \textit{deathplace} we observed during the evaluation is confirmed here. In addition, recall drops significantly for the \textit{Other} class, mainly due to the fact that this was increased because of incorrect classifications by the matching algorithm.

\begin{table*}[!ht]
\begin{tabular}{|c|lll|lll|lll|}
\hline
\multirow{2}{*}{} & \multicolumn{3}{c|}{\textbf{normal}} & \multicolumn{3}{c|}{\textbf{coref}} & \multicolumn{3}{c|}{\textbf{skip}} \\ \cline{2-10} 
 & \multicolumn{1}{c}{P} & \multicolumn{1}{c}{R} & \multicolumn{1}{c|}{F1} & \multicolumn{1}{c}{P} & \multicolumn{1}{c}{R} & \multicolumn{1}{c|}{F1} & \multicolumn{1}{c}{P} & \multicolumn{1}{c}{R} & \multicolumn{1}{c|}{F1} \\ \hline
birthdate & 1.0 & 1.0 & 1.0 & 0.99 & 1.0 & 0.99 & 1.0 & 1.0 & 1.0 \\
birthplace & 0.84 & 0.9 & 0.87 & 0.86 & 0.88 & 0.87 & 0.79 & 0.83 & 0.81 \\
deathdate & 1.0 & 0.99 & 1.0 & 0.98 & 1.0 & 0.99 & 0.94 & 0.99 & 0.96 \\
deathplace & 0.37 & 0.95 & 0.53 & 0.31 & 1.0 & 0.48 & 0.36 & 0.97 & 0.53 \\
occupation & 0.8 & 1.0 & 0.89 & 0.9 & 1.0 & 0.85 & 0.68 & 1.0 & 0.81 \\
educatedAt & 0.88 & 1.0 & 0.94 & 0.92 & 0.99 & 0.95 & 0.96 & 0.99 & 0.97 \\
ofParent & 0.77 & 0.99 & 0.87 & 0.73 & 1.0 & 0.85 & 0.8 & 1.0 & 0.89 \\
hasChild & 0.8 & 0.99 & 0.88 & 0.64 & 1.0 & 0.78 & 0.63 & 1.0 & 0.77 \\
sibling & 0.75 & 0.95 & 0.84 & 0.62 & 1.0 & 0.77 & 0.7 & 0.92 & 0.8 \\
other & 0.97 & 0.37 & 0.54 & 0.98 & 0.36 & 0.53 & 0.96 & 0.33 & 0.49 \\ \hline
macro avg. & 0.82 & 0.91 & 0.83 & 0.79 & 0.92 & 0.81 & 0.78 & 0.9 & 0.8 \\ \hline
\end{tabular}
\caption{Evaluation of Manual Annotations on Gold Set}
\label{tab:gold_eval}
\end{table*}

\subsection{Neural Models}
\label{sec:neural_eval}

The machine learning model we used to perform relationship classification is based on transformers. Since their introduction, transformer models have shown excellent results in various NLP tasks \cite{devlin2019bert} such as text classification \cite{ranasinghe-zampieri-2020-multilingual}, NER \cite{jia-etal-2020-entity} and question answering \cite{yang-etal-2019-end-end} including RE \cite{yamada-etal-2020-luke,joshi-etal-2020-spanbert,10.1145/3357384.3358119,alt2019improving}. In this research, we utilised the architecture introduced by \citet{baldini-soares-etal-2019-matching}. 

\begin{figure}[!ht]
\centering
\scalebox{0.3}{\includegraphics{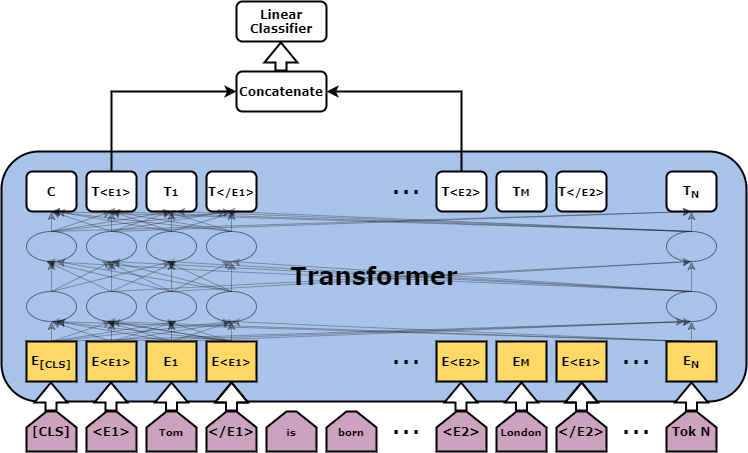}}
\caption{Neural Network Architecture}
\label{fig:neural_architecture}
\end{figure}

The input to the transformer models is the sentence with “[E1]” and “[E2]” markers marking the positions of their respective entities. Then the output hidden states of transformer at the “[E1]” and “[E2]” token positions are concatenated as the final output representation of the relationship. Finally, a linear classifier is stacked on top of the output representation. The architecture diagram is visualised in Figure \ref{fig:neural_architecture}.

We fine-tune all the parameters from the transformer as well as the linear classifier jointly by maximising the log-probability of the correct label. For all the experiments we optimised parameters (with AdamW) using a learning rate of $7e-5$, a maximum sequence length of $512$, and a batch size of $32$ samples. The models were trained using a $24$ GB RTX 3090 GPU over five epochs. As the pre-trained transformer model, we used the \textit{bert-base-uncased} model available in HuggingFace \cite{wolf-etal-2020-transformers}. 

For training the BERT-based classifier, we used each of the three sets separately, as well as a combination of the three sets we refer to as \textit{all}, where we remove any duplicates that might be caused by the combination. We did not focus on producing the best possible results, and rather on indicating whether the produced dataset is even suitable for training a model. Table \ref{tab:models_eval} shows the evaluation results of the models trained on the four different sets. While largely the results of the matching algorithm are echoed, we observe that some other relations, including \textit{hasChild, ofParent} and \textit{sibling} seem to score quite low in terms of recall. When comparing to the counts per set (See Table \ref{tab:no_rels_improv}) these relations are quite low in number compared to the others, possibly explaining the results.

\begin{table*}[!ht]
\begin{tabular}{|c|lll|lll|lll|lll|}
\hline
\multirow{2}{*}{} & \multicolumn{3}{c|}{\textbf{normal}} & \multicolumn{3}{c|}{\textbf{coref}} & \multicolumn{3}{c|}{\textbf{skip}} & \multicolumn{3}{c|}{\textbf{all}} \\ \cline{2-13} 
                  & \multicolumn{1}{c}{P} & \multicolumn{1}{c}{R} & \multicolumn{1}{c|}{F1} & \multicolumn{1}{c}{P} & \multicolumn{1}{c}{R} & \multicolumn{1}{c|}{F1} & \multicolumn{1}{c}{P} & \multicolumn{1}{c}{R} & \multicolumn{1}{c|}{F1} & \multicolumn{1}{c}{P} & \multicolumn{1}{c}{R} & \multicolumn{1}{c|}{F1} \\ \hline
birthdate & 1.0 & 0.99 & 1.0 & 1.0 & 0.99 & 0.99 & 0.87 & 0.92 & 0.89 & 1 & 0.99 & 0.99 \\
birthplace & 0.85 & 0.77 & 0.81 & 0.8 & 0.8 & 0.8 & 0.81 & 0.74 & 0.8 & 0.88 & 0.8 & 0.83 \\
deathdate & 1.0 & 0.95 & 0.97 & 1.0 & 0.98 & 0.99 & 0.98 & 0.86 & 0.91 & 1.0 & 0.98 & 0.99 \\
deathplace & 0.73 & 0.53 & 0.62 & 0.79 & 0.52 & 0.62 & 0.82 & 0.42 & 0.55 & 0.81 & 0.48 & 0.6 \\
occupation & 1.0 & 0.99 & 1.0 & 1.0 & 1.0 & 1.0 & 1.0 & 0.98 & 0.99 & 1.0 & 0.99 & 1.00 \\
educatedAt & 0.98 & 0.87 & 0.92 & 0.97 & 0.91 & 0.94 & 0.98 & 0.83 & 0.9 & 1.0 & 0.87 & 0.93 \\
ofParent & 0.92 & 0.54 & 0.66 & 0.82 & 0.57 & 0.67 & 0.78 & 0.51 & 0.61 & 0.88 & 0.6 & 0.7 \\
hasChild & 0.96 & 0.36 & 0.42 & 0.92 & 0.43 & 0.55 & 0.97 & 0.39 & 0.5 & 0.98 & 0.39 & 0.49 \\
sibling & 0.92 & 0.45 & 0.57 & 0.94 & 0.43 & 0.55 & 0.87 & 0.45 & 0.55 & 0.93 & 0.46 & 0.57 \\
other & 0.38 & 0.95 & 0.54 & 0.41 & 0.94 & 0.57 & 0.38 & 0.93 & 0.54 & 0.39 & 0.95 & 0.56 \\ \hline
macro avg. & 0.9 & 0.73 & 0.76 & 0.89 & 0.75 & 0.78 & 0.87 & 0.7 & 0.74 & 0.92 & 0.74 & 0.78 \\ \hline
\end{tabular}
\caption{Evaluation Metrics for Relations in each Set}
\label{tab:models_eval}
\end{table*}

\section{Proposed Application} 
\label{sec:discussion}
% Section for Spencer explaining how and why he would use it for historical research
The availability of compiled datasets for historical research is more important than ever. While NLP methods in domains such as biomedical and news continue to be expanded greatly, smaller areas of research such as specific historical (biographical) research inherently lack these opportunities. Being able to compile datasets to train neural extraction models with relative ease, as described here, is crucial for future research. In this section, we highlight this by example of the Army List in the United Kingdom, a study that we plan to embark on in the coming months.

The Army List \cite{1913Tqal} is a biographical compendium of officers serving in the British Army. It was first published in 1840 and volumes were subsequently published annually, although this varied during wartime. Each volume lists the name and rank of every serving officer in the British Army, along with important biographical details including length of service, past roles, and current position held. The Army List is an essential starting point for any research about the careers of military officers in the period. 

Despite its importance to historical research, the Army List can prove difficult to access. Copies are held by a handful of specialist archives in the United Kingdom and there has been no systematic attempt to digitise them or apply data processing to them. 
Each Army List contains a wealth of information that invites cross-referencing and comparison to learn more about professional and social links amongst the officer class. However, the sheer number of biographical entries, amounting to several thousand per volume, made this an impossible task for historians in the pre-digital age. Digital processing offers a solution to this problem and opens the possibility of being able to map connections in new and illuminating ways. For example, it would allow the identification of professional networks based on age, shared roles, unit associations, and overseas service. A dataset based upon it would be of enormous value to historians, and it would open exciting new avenues for research and would contribute to ongoing historiographical debates on the professional bonds of the officer class.

To enable the kind of research described above, there is clearly a need for datasets like \textit{Biographical} so that systems can be trained to extract large amounts of information quickly and efficiently. Not only could the dataset we present here be used itself, but also new datasets, compiled with the method we present here. Both the dataset and method therefore present significant opportunities for application, enabling research in under-resourced areas.

\section{Conclusion}
We have presented \textit{Biographical}, a relation extraction dataset that is semi-supervised, and described its compilation process in detail. Furthermore, we carried out a number of experiments to understand the dataset better. This included different processing approaches, a manual annotation task and the training of different neural models. Not only have these experiments investigated different ways of optimising the compilation of the dataset for different goals, they have also validated the results in terms of machine learning.

In more general terms, this work marks an exciting first step at applying data processing to historical documentation. Archival digitisation in the United Kingdom and other countries remains hesitant and inconsistent, and there has been very little data processing of that which is available. The application of more computational resources to mine the data would be of immense value to historians and those working in related fields. 

In the future, we would like to address a number of different aspects concerning this dataset. First, the optimisation of the compilation process for even more precise results will be focused on. Next, we would like to extend the number of relations, and demonstrate how simple this could be. As mentioned in the previous section, we also intend to test this approach on real-world texts in collaboration with historians.

\bibliographystyle{ACM-Reference-Format}
\bibliography{references}


\end{document}